\documentclass{aa}
\usepackage{amssymb}
\usepackage{amsmath}
\usepackage{xspace}
\usepackage{graphicx}
\usepackage{enumerate}   
\usepackage{color,xcolor}
\usepackage{txfonts}
\def\gr{$\gamma$-ray}

\usepackage{natbib}
\usepackage[colorlinks=true,citecolor=blue,urlcolor=blue,linkcolor=blue]{hyperref}

\def\gr{$\gamma$-ray}

\begin{document}

\title{Revision of upper bound on volume-filling intergalactic magnetic fields with LOFAR}

\author{A. Neronov\inst{1,2}, F. Vazza\inst{3,4}, S. Mtchedlidze\inst{3,5}, E. Carretti\inst{4}}
\institute{BAstroparticules et Cosmologie, Universit\'e Paris Cit\'e, CNRS, France
  \and Laboratory of Astrophysics, \'Ecole Polytechnique F\'ed\'erale de Lausanne, 1015 Lausanne, Switzerland
  \and Dipartimento di Fisica e Astronomia, Universita di Bologna, Via Gobetti 93/2, 40129 Bologna, Italy 
  \and INAF Istituto di Radioastronomia, Via P. Gobetti 101, 40129 Bologna, Italy
  \and School of Natural Sciences and Medicine, Ilia State University, 3-5 Cholokashvili St., 0194 Tbilisi, Georgia}

\authorrunning{Neronov et al.}
\titlerunning{}
\abstract
{
Magnetic fields present in the Large Scale Structure (LSS) of the Universe change polarization of radio waves arriving from distant extragalactic sources through the effect of Faraday rotation. This effect has been recently used to detect magnetic field in the LSS filaments based on the Rotation Measure data of the LOFAR Two-Meter Sky Survey (LoTSS). We notice that the same data also constrain the strength of the volume-filling magnetic field in the voids of the LSS. We use the LoTSS  data to to derive an improved upper bound on the volume-filling field. The new upper bound provides an order of magnitude improvement on the previous Faraday rotation bounds. The new Faraday Rotation bound on the scale-invariant field  that may originate from the epoch of inflation is also an order of magnitude lower than the bound on such field derived from the anisotropy analysis of the Cosmic Microwave Background. 
}

\keywords{}
\maketitle

\section{Introduction}

The existence of non-zero magnetic fields in the voids of the large-scale structure has been hinted at by non-observation of extended and delayed secondary \gr ~ emission from distant extragalactic sources, such as Active Galactic Nuclei (AGN)   \cite{2010Sci...328...73N,MAGIC:2022piy,HESS:2023zwb} and Gamma-Ray Bursts (GRB) \cite{Vovk:2023qfk}.
%
Magnetic fields in voids can be of cosmological (primordial) origin.
As proposed by inflationary magnetogenesis theories, primordial magnetic fields (PMFs) correlated on Mpc scales are generated when the conformal invariance of the electromagnetic action is broken via the coupling of the electromagnetic field with the scalar inflaton field during inflation (see Refs. \cite{TurnerWidrow1988,Ratra1992} for pioneering studies and Ref. \cite{Durrer:2013pga} for a review). 
PMFs could have also been produced during cosmic phase transitions through charge separation processes (see Refs.~\cite{Hogan1983,Quashnocketal1989,Tajimaetal1992} and Refs.~\cite{Subramanian2016,Vachaspati:2020blt} for reviews). In this case, their coherence scale $\lambda_{\rm{B}}$, is initially smaller than the Hubble horizon at the electroweak (EW) or quantum chromodynamical (QCD) phase-transition epochs, although it can reach tens or hundreds of kpc scales (see e.g., Refs.~\cite{Tevzadzeetal2012,Brandenburgetal2017,Durrer:2013pga}) by the end of recombination
when accounting for the magnetohydrodynamic (MHD) turbulent decay (i.e., the inverse cascade process -- decrease of the magnetic field strength $B$, accompanied by an increase of $\lambda_{\rm{B}}$) of the field after its generation; the strength and coherence scale of the magnetic field at the recombination, and consequently, at the current epoch depend on the decay timescale of the field and $B$--$\lambda_B$ evolutionary trend \cite{Banerjee:2004df,Brandenburgetal2017,Hosking:2022umv,Brandenburg:2024tyi}. 
Alternatively, magnetic fields could have been generated and amplified during structure formation processes, and then transported into voids through magnetised outflows from galaxies \cite{Bertone:2006mr,Donnertetal2009} or AGN jets \cite{Xuetal2011}.
Even though the most recent modeling of this process indicate that  galactic outflows are perhaps not powerful enough for producing magnetised outflows that result in the void volume-filling magnetic fields \cite{Marinacci:2017wew,va17cqg, Bondarenko:2021fnn,Blunier:2024aqx,Carretti:2024bcf}, modeling uncertainties are large. 

Various observational probes have been used to place constraints on the void Inter-Galactic Magnetic Field (IGMF) strength. 
Gamma-ray observations yield a lower bound on the void IGMF  strength $B$ that depends on assumptions about the field variability distance scale $\lambda_B$ \cite{2010Sci...328...73N,MAGIC:2022piy,HESS:2023zwb}. For the large scale magnetic field variable on the Hubble distance scale, the most conservative lower bound on the field strength is currently at the level of $\sim 10^{-17}$~G \cite{MAGIC:2022piy}. Upper limits have been previously derived from non-detection of Faraday Rotation of the polarized radio waves from AGN \cite{Blasi:1999hu,Pshirkov:2015tua,Aramburo-Garcia:2022ywn}. This upper limit also depends on the magnetic field spatial structure and for the large-scale field it is at the level of several nano-Gauss \cite{Aramburo-Garcia:2022ywn}. A correlation-length dependent upper bound can also be deduced from Ultra-High-Energy Cosmic Ray observations \cite{Neronov:2021xua}. If the field is of cosmological origin, it is also constrained by non-detection of magnetic field induced anisotropies on the Cosmic Microwave Background (CMB) \cite{Planck:2015zrl,Zucca:2016iur,2019JCAP...11..028P}.

Faraday Rotation measurements, in particular, the Residual Rotation Measure (RRM\footnote{The RRM refers to the Rotation Measure (RM) line-of-sight integral $RM\propto \int n_e B_{||}dl$ of the product of the free electron density $n_e$ and the magnetic field component parallel to the line-of-sight, $B_{||}$ that is obtained after subtraction of the modeled  contribution of the contribution of the Milky Way galaxy to the integral.}) analysis, have recently been used to study the effects of PMF spectrum \cite{Carrettietal2023,Carretti:2024bcf,Mtchedlidze:2024kvt} and coherence scale \cite{Mtchedlidze:2024kvt} on the redshift evolution of the RRM. 
In what follows,
we use the simulated RRM from Refs.~\cite{Carretti:2024bcf,Mtchedlidze:2024kvt} and the constrained RRM \cite{Carretti:2024bcf}, derived from observations of the Faraday rotation measure of polarized extragalactic sources up to redshift $z \approx 2.5$ (from the LOFAR Two-metre
Sky Survey, LoTSS, \cite{2023MNRAS.519.5723O}), to place upper limits on the  void IGMF strength.
%
%
%
We show that the new LOFAR bound is an order of magnitude tighter than the bound from the CMB anisotropy studies \cite{Planck:2015zrl,Zucca:2016iur}, particularly for inflationary PMF with scale-invariant power spectrum,
 while it is comparable to the CMB anisotropy bound for the cosmological fields that may originate from cosmological phase transitions. 

\section{Upper limit from LOFAR RRM}

The RRM redshift-dependence data, RRM($z$), were obtained in Ref.~\cite{Carretti:2024bcf} by analysing the LoTSS data, after subtraction of the Galactic foreground, $\mathrm{RM}_{Gal}$, to the total RM, $\mathrm{RM}_{total}$ ($\mathrm{RRM} = \mathrm{RM}_{total} - \mathrm{RM}_{Gal}$) and removing the RM contribution from intervening massive halos along the lines of sight.
%
The reconstructed RRM($z$) trends 
exhibit significant 
scatter of unclear origin (including ``wiggles''  across all redshift  bins, likely with a physical origin.
In this work, we follow a conservative approach, in which we consider the 
RRM($z$) trend from Ref.~\cite{Carretti:2024bcf}
%
as an upper bound on the contribution from IGMF, noticing that 
the RM induced by the IGMF cannot exceed the RRM. 
%
 For each 
RRM model simulated using cosmological simulations with the ENZO code, reported in Ref.~\cite{Carretti:2024bcf}, we find the best-fit to the {constrained} RRM data and define the maximal possible magnetic field strength as the value at which the model becomes inconsistent with the data at 90\% confidence level.  In this way, we derive the maximum possible normalization, $A$, of the magnetic field power spectrum, defined in Fourier space $P(k)=A(k/k_0)^{\alpha}$ for different slopes $\alpha$. We subsequently re-express the magnetic field power spectrum in real space as $B(\lambda,\alpha)$, 
representing the magnetic field strength
averaged over given distance scale $\lambda$. In fact, Ref.~\cite{Carretti:2024bcf} directly provides normalization of this real-space power spectrum as the magnetic field strength $B_{\rm{Mpc}}$ averaged over the reference scale $1$~Mpc, following the convention used in the analysis of the CMB (see e.g., Ref.~\cite{Planck:2015zrl}).

 \begin{figure}
     \includegraphics[width=\columnwidth]{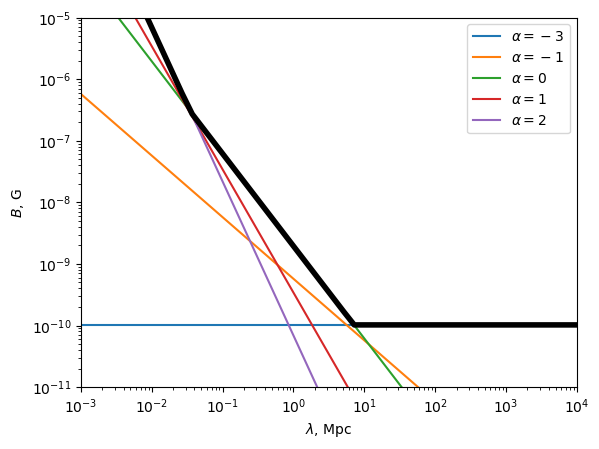}
     \caption{Upper bounds on the field as a function of distance scale for {PMF} models 
     where magnetic field is characterised by the power-law power spectrum with different slopes (colored lines).
     Black line shows the envelope of all colored lines, which is an upper bound on the field strength marginalized over the slope $\alpha$. }
     \label{fig:envelope}
 \end{figure}
 
 Fig.~\ref{fig:envelope} shows the upper bounds on the field strength as a function of distance scale for different assumptions about the slope of the power spectrum $\alpha$ and the overall upper bound marginalized over $\alpha$. 
The magnetic field energy density is related to the power spectrum as $E_M=\int 4\pi k^2P(k)dk$. The minimal possible value of $\alpha=-3$ corresponds to the scale-invariant magnetic field (which only inflationary processes can produce) that has magnetic field energy density per decade of $k$ independent of $k$. Such magnetic field is represented by horizontal line part of the upper bound shown in Fig.~\ref{fig:envelope}. The largest value of $\alpha$ explored in Ref.~\cite{Carretti:2024bcf} is $\alpha=2$, corresponding to the 
PMFs produced during or after reheating.
The upper limit on 
{the PMF}
with such power spectrum is represented by the steepest part of the upper limit curve in Fig.~\ref{fig:envelope}. The field with such power spectrum is represented by a straight line $B=B_{\rm{Mpc}} (\lambda_B/1\mbox{ Mpc})^{-(\alpha+3)/2}=B_{\rm{Mpc}} (\lambda_B/1\mbox{ Mpc})^{-5/2}$ in the log-log plot of the figure. The line corresponding to the maximal possible field of this type is tangent to the upper limit black curve in the figure.

It should be noted that any PMF model is affected by a small-scale damping on scales smaller than the Silk damping scale for acoustic waves in cosmological perturbations, caused 
alternating phases of turbulent and viscous regimes (see e.g., Ref.~\cite{Banerjee:2004df}).
This means that, for every particular magnetic field configuration, there is a minimum scale, $l_D$, below which the field has no structure and our $(B,\lambda)$ relation cannot be extrapolated for  $\lambda \leq l_D$ scales. 
The physics ruling the exact value of $l_D$ is complex (see e.g., Refs.~\cite{Trivedi18,Jedamzik_Abel_23}), and it depends on the combination of the spectral energy distribution of magnetic fields, as well as on the normalisation of the field. However, for the range of values in which we are concerned here, 
$l_D$ is usually assumed to be in the $\sim 10-100$~kpc range.
(see e.g., \cite{2019JCAP...11..028P}). Dissipation of the small-scale power results 
in the formation of a broken power-law type power spectra,  with a steeper slope above the $k=2\pi/l_D$ wavenumber.

\begin{figure}
\includegraphics[width=\columnwidth]{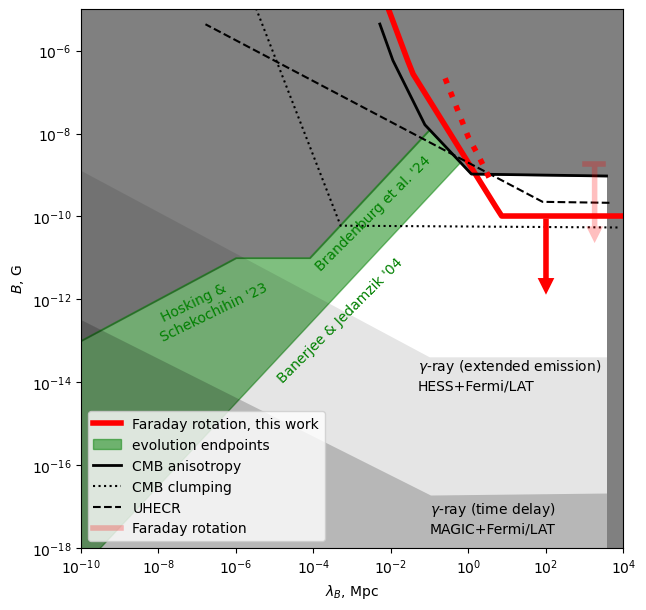}
\caption{LOFAR RM bounds on IGMF strength and correlation length, compared with other known constraints. Green area shows possible locations of endpoints of the evolution of PMF \cite{Hosking:2022umv,Brandenburg:2024tyi,Banerjee:2004df} {generated during EW or QCD phase transitions [?].}. The red semi-transparent arrow shows the level of other recent limits from the Faraday rotation, obtained by comparing Illustris simulations with NVSS
radio data \cite{Aramburo-Garcia:2022ywn}. \gr\ lower bounds are from Refs. \cite{MAGIC:2022piy,HESS:2023zwb}.}
\label{fig:bounds}
\end{figure}

For {power-law} magnetic fields with a slope $\alpha>-2$, 
{or for magnetic fields with a more complex power spectrum than a simple power-law (e.g., PMF models explored in Refs.~\cite{Brandenburgetal2017,Mtchedlidze:2024kvt}),}
it is possible to define the magnetic field correlation length $l_B=\int k^{-1}P(k)d^3k/E_M$. 
Different estimates of the comoving magnetic field correlation length at recombination have been made in the past,
and recently. 
{As mentioned in the Introduction the locus of PMF states in the $B - \lambda_{B}$ parameter space depends on the magnetic field decay timescale}. 
%
The green band in Fig.~\ref{fig:bounds} illustrates the uncertainty in the predictions of the magnetic field strength and coherence scale.
An order-of-magnitude estimate for $B$ and $\lambda_{B}$ can be obtained by {assuming that pre-recobmination decay of the field is governed by Alfv\'enic timescale and that it is 
of the order of
Hubble time.} 
{That is,} $l_B=C_M^{-1} v_At_H$, where {$v_A$ is Alfv\'en velocity, $t_H$ is the Hubble time and,} $C_M$ is a proportionality coefficient. This estimate, assuming $C_M\simeq 1$ \cite{Banerjee:2004df} is shown with the ``Banerjee \& Jedamzik '04'' label in the figure. It has been argued in Ref.~\cite{Hosking:2022umv} that reconnection of magnetic field lines plays an important role in regulating the rate of turbulent decay of small-scale eddies, and $C_M$ can be much larger than unity. These estimates from Ref.~\cite{Hosking:2022umv} are shown at the boundary of the green region labeled ``Hosking \& Schekochihin '23'' in Fig.~\ref{fig:bounds}. The direct numerical modeling of Ref.~\cite{Brandenburg:2024tyi} indeed reveals the values of $C_M\gg 1$, reaching $C_M\simeq 50$, but it does not confirm the hypothesis proposed in Ref.~\cite{Hosking:2022umv} on the strong dependence of $C_M$ on the Prandtl number (the ratio of viscosity and resistivity of the plasma). The estimates from Ref.~\cite{Brandenburg:2024tyi} are shown  within the green region  labeled ``Brandenburg et al. '24''.

To explore the dependence of the upper bound from the constrained RRM on the $l_B$ assumptions, we use numerical simulations from Ref.~\cite{Mtchedlidze:2024kvt}. These simulations consider PMF models with correlation lengths ($ l_B=\left\{1\mbox{ Mpc, } 1.8\mbox{ Mpc, }  3.5\mbox{ Mpc}\right\}$ as initial conditions for the ENZO cosmological MHD simulations for the  magnetic fields with broken powerlaw spectra with the slopes  $\alpha=2$ at small $k$ and $\alpha=-11/3$ in the large $k$ limit. While these models may resemble phase-transition-generated PMFs, the aforementioned coherence scales are larger than those currently predicted by theory. For these PMF models (see Table~1 in Ref.~\cite{Mtchedlidze:2024kvt}), we find the scaling factor of magnetic field power spectrum at which the predicted RM would saturate the LOFAR upper bound. In this way we obtain the upper bounds on $B$ as a function of $l_B$. We find that these upper bounds follow a power-law  shown by the red dotted line in Fig.~\ref{fig:bounds}. The bound shown by the red dashed line is a slight improvement of the bound derived in Ref.~\cite{Mtchedlidze:2024kvt} due to the update of the LOFAR RRM data in Ref.~\cite{Carretti:2024bcf}. It should be stressed that the bounds represented by the red dotted line and by the red solid curve have somewhat different meaning: 
\begin{itemize}
    \item the dotted red line is the bound on $B$ -- $l_B$ for magnetic fields with ``causal'' power spectra with a $\alpha=2$ slope for small $k$;
    \item the solid red line is the bound on magnetic fields with different spectral slopes $\alpha$ ($\alpha\not= 2$ corresponds to magnetic fields produced in ``acausal'' way, i.e. originating from inflation). 
\end{itemize}

\section{Discussion}

Fig.~\ref{fig:bounds} shows that the new LOFAR upper limit on the volume-filling IGMF is particularly strong for the fields with scale-invariant power spectrum (that would be represented by horizontal lines in the figure). The normalization of such field at the reference smoothing scale $\lambda=1$~Mpc cannot exceed $B_{\rm{Mpc}}=70$~pG. This is more than an order of magnitude below the limit that has been derived from the analysis of CMB anisotropies \cite{Planck:2015zrl,Zucca:2016iur}. The influence of magnetic fields on cosmological recombination is not limited to the CMB anisotropy. The coupling of the magnetic field to the primordial plasma creates inhomogeneities in the baryonic fluid, which modify the dynamics of recombination. It has been argued that this effect is observable, and
when accounted for, it
leads to a modification of the expansion rate of the Universe {as inferred} from the CMB data \cite{Jedamzik:2020krr}. Such modification is relaxing the so-called ``Hubble tension'' problem of cosmology --- an inconsistency between the estimates of the expansion rate of the Universe derived from the measurements in the present-day Universe and from cosmological probes in the earlier epochs. 

In the absence of unambiguous detection of the effect, an upper limit on the present-day relic magnetic field in the voids have been derived in Ref.~\cite{Jedamzik:2018itu}. This limit is shown by the dotted black line in Fig.~\ref{fig:bounds}. Remarkably, this bound is close to the new Faraday Rotation bound derived above. This means that if 
the 
PMF
is indeed relevant for the solution of the Hubble tension problem, and the field is a scale-invariant originating from inflation ({see also footnote on p.2}), it should be detectable through its effect on the Faraday rotation of polarized radio signals. 
Major improvements of the Faraday rotation data, expected with the next releases of LOFAR, ASKAP and SKA, may ultimately lead to the detection of the volume-filling inflationary IGMF in this case.

Otherwise, for the causally produced cosmological magnetic fields, or inflationary fields with the steeper spectral slopes $\alpha>-3$,  LOFAR constraints are currently weaker than the
constraints obtained through
CMB anisotropy and are much weaker than the CMB clumping constraints. Prospects for detection of such fields in the Faraday rotation data are therefore not promising. The detection of the signature of such shorter correlation length fields using the Faraday Rotation technique would have several important consequences.
First, it may imply that the volume filling fields are non-cosmological (since they would have not impacted CMB observables), but it would also strongly challenge all commonly proposed alternative scenarios for cosmic magnetisation from galactic outflows or jets from active galactic nuclei (see e.g., Refs.~\cite{Bertone:2006mr,Marinacci:2017wew,va17cqg,Bondarenko:2021fnn}) since magnetic fields generated in this way must still have $\sim  \rm Mpc$ correlation lengths (see e.g., Ref.~\cite{va17cqg}). Alternatively, this will imply that the complex physics of the interplay between PMFs and baryons in the recombination era \cite{Trivedi18,2019MNRAS.484..185P,Jedamzik_Abel_23} requires further assessment. 

In conclusion, we surmise that we might have entered an interesting era, in which the combination of radio and \gr\ observations of the low redshift Universe, combined with the advanced theoretical modeling of the co-evolution of magnetic fields with cosmic structures, can allow us to probe (or limit) PMFs better than what is possible using the existing or 
forthcoming
CMB data, for a significant range of magnetic field configurations.

\section*{Acknowledgments}
The authors acknowledge the COsmic Magnetism with RADio Astronomy 2024 (COMRAD2024) conference for fostering insightful discussions that contributed to the development of some of the ideas presented in this work. F.V. and S. M.'s research has been supported by Fondazione Cariplo and Fondazione CDP, through grant  Rif: 2022-2088 CUP J33C22004310003 for "BREAKTHRU" project. F.V. acknowledges the CINECA award  "IsB28\_RADGALEO" under the ISCRA initiative, for the availability of high-performance computing resources and support used to produce the numerical models used in this work. A.N. has been supported in part by the French National Research Agency (ANR) grant ANR-24-CE31-4686

\bibliographystyle{aa}
\bibliography{refs.bib}

\begin{thebibliography}{40}
\expandafter\ifx\csname natexlab\endcsname\relax\def\natexlab#1{#1}\fi

\bibitem[{Acciari {et~al.}(2023)}]{MAGIC:2022piy}
Acciari, V.~A. {et~al.} 2023, Astron. Astrophys., 670, A145

\bibitem[{Ade {et~al.}(2016)}]{Planck:2015zrl}
Ade, P. A.~R. {et~al.} 2016, Astron. Astrophys., 594, A19

\bibitem[{Aharonian {et~al.}(2023)}]{HESS:2023zwb}
Aharonian, F. {et~al.} 2023, Astrophys. J. Lett., 950, L16

\bibitem[{Aramburo-Garcia {et~al.}(2022)Aramburo-Garcia, Bondarenko, Boyarsky,
  Neronov, Scaife, \& Sokolenko}]{Aramburo-Garcia:2022ywn}
Aramburo-Garcia, A., Bondarenko, K., Boyarsky, A., {et~al.} 2022, Mon. Not.
  Roy. Astron. Soc., 515, 5673

\bibitem[{Banerjee \& Jedamzik(2004)}]{Banerjee:2004df}
Banerjee, R. \& Jedamzik, K. 2004, Phys. Rev. D, 70, 123003

\bibitem[{Bertone {et~al.}(2006)Bertone, Vogt, \& Ensslin}]{Bertone:2006mr}
Bertone, S., Vogt, C., \& Ensslin, T. 2006, Mon. Not. Roy. Astron. Soc., 370,
  319

\bibitem[{Blasi {et~al.}(1999)Blasi, Burles, \& Olinto}]{Blasi:1999hu}
Blasi, P., Burles, S., \& Olinto, A.~V. 1999, Astrophys. J. Lett., 514, L79

\bibitem[{Blunier \& Neronov(2024)}]{Blunier:2024aqx}
Blunier, J. \& Neronov, A. 2024, Astron. Astrophys., 691, A34

\bibitem[{Bondarenko {et~al.}(2022)Bondarenko, Boyarsky, Korochkin, Neronov,
  Semikoz, \& Sokolenko}]{Bondarenko:2021fnn}
Bondarenko, K., Boyarsky, A., Korochkin, A., {et~al.} 2022, Astron. Astrophys.,
  660, A80

\bibitem[{{Brandenburg} {et~al.}(2017){Brandenburg}, {Kahniashvili}, {Mandal},
  {Pol}, {Tevzadze}, \& {Vachaspati}}]{Brandenburgetal2017}
{Brandenburg}, A., {Kahniashvili}, T., {Mandal}, S., {et~al.} 2017, \prd, 96,
  123528

\bibitem[{Brandenburg {et~al.}(2024)Brandenburg, Neronov, \&
  Vazza}]{Brandenburg:2024tyi}
Brandenburg, A., Neronov, A., \& Vazza, F. 2024, Astron. Astrophys., 687, A186

\bibitem[{{Carretti} {et~al.}(2023){Carretti}, {O'Sullivan}, {Vacca}, {Vazza},
  {Gheller}, {Vernstrom}, \& {Bonafede}}]{Carrettietal2023}
{Carretti}, E., {O'Sullivan}, S.~P., {Vacca}, V., {et~al.} 2023, \mnras, 518,
  2273

\bibitem[{Carretti {et~al.}(2024)Carretti, Vazza, O'Sullivan, Vacca, Bonafede,
  Heald, Horellou, Mtchedlidze, \& Vernstrom}]{Carretti:2024bcf}
Carretti, E., Vazza, F., O'Sullivan, S.~P., {et~al.} 2024
  [\eprint[arXiv]{2411.13499}]

\bibitem[{{Donnert} {et~al.}(2009){Donnert}, {Dolag}, {Lesch}, \&
  {M{\"u}ller}}]{Donnertetal2009}
{Donnert}, J., {Dolag}, K., {Lesch}, H., \& {M{\"u}ller}, E. 2009, Mon. Not.
  Roy. Astron. Soc., 392, 1008

\bibitem[{Durrer \& Neronov(2013)}]{Durrer:2013pga}
Durrer, R. \& Neronov, A. 2013, Astron. Astrophys. Rev., 21, 62

\bibitem[{{Hogan}(1983)}]{Hogan1983}
{Hogan}, C.~J. 1983, \prl, 51, 1488

\bibitem[{Hosking \& Schekochihin(2023)}]{Hosking:2022umv}
Hosking, D.~N. \& Schekochihin, A.~A. 2023, Nature Commun., 14, 7523

\bibitem[{{Jedamzik} {et~al.}(2023){Jedamzik}, {Abel}, \&
  {Ali-Haimoud}}]{Jedamzik_Abel_23}
{Jedamzik}, K., {Abel}, T., \& {Ali-Haimoud}, Y. 2023, arXiv e-prints,
  arXiv:2312.11448

\bibitem[{Jedamzik \& Pogosian(2020)}]{Jedamzik:2020krr}
Jedamzik, K. \& Pogosian, L. 2020, Phys. Rev. Lett., 125, 181302

\bibitem[{Jedamzik \& Saveliev(2019)}]{Jedamzik:2018itu}
Jedamzik, K. \& Saveliev, A. 2019, Phys. Rev. Lett., 123, 021301

\bibitem[{Marinacci {et~al.}(2018)}]{Marinacci:2017wew}
Marinacci, F. {et~al.} 2018, Mon. Not. Roy. Astron. Soc., 480, 5113

\bibitem[{Mtchedlidze {et~al.}(2024)Mtchedlidze, Domínguez-Fernández, Du,
  Carretti, Vazza, O’Sullivan, Brandenburg, \&
  Kahniashvili}]{Mtchedlidze:2024kvt}
Mtchedlidze, S., Domínguez-Fernández, P., Du, X., {et~al.} 2024, The
  Astrophysical Journal, 977, 128

\bibitem[{Neronov {et~al.}(2023)Neronov, Semikoz, \&
  Kalashev}]{Neronov:2021xua}
Neronov, A., Semikoz, D., \& Kalashev, O. 2023, Phys. Rev. D, 108, 103008

\bibitem[{{Neronov} \& {Vovk}(2010)}]{2010Sci...328...73N}
{Neronov}, A. \& {Vovk}, I. 2010, Science, 328, 73

\bibitem[{{O'Sullivan} {et~al.}(2023){O'Sullivan}, {Shimwell}, {Hardcastle},
  {Tasse}, {Heald}, {Carretti}, {Br{\"u}ggen}, {Vacca}, {Sobey}, {Van Eck},
  {Horellou}, {Beck}, {Bilicki}, {Bourke}, {Botteon}, {Croston}, {Drabent},
  {Duncan}, {Heesen}, {Ideguchi}, {Kirwan}, {Lawlor}, {Mingo},
  {Nikiel-Wroczy{\'n}ski}, {Piotrowska}, {Scaife}, \& {van
  Weeren}}]{2023MNRAS.519.5723O}
{O'Sullivan}, S.~P., {Shimwell}, T.~W., {Hardcastle}, M.~J., {et~al.} 2023,
  \mnras, 519, 5723

\bibitem[{{Paoletti} {et~al.}(2019){Paoletti}, {Chluba}, {Finelli}, \&
  {Rubi{\~n}o-Mart{\'\i}n}}]{2019MNRAS.484..185P}
{Paoletti}, D., {Chluba}, J., {Finelli}, F., \& {Rubi{\~n}o-Mart{\'\i}n}, J.~A.
  2019, \mnras, 484, 185

\bibitem[{{Paoletti} \& {Finelli}(2019)}]{2019JCAP...11..028P}
{Paoletti}, D. \& {Finelli}, F. 2019, \jcap, 2019, 028

\bibitem[{Pshirkov {et~al.}(2016)Pshirkov, Tinyakov, \&
  Urban}]{Pshirkov:2015tua}
Pshirkov, M.~S., Tinyakov, P.~G., \& Urban, F.~R. 2016, Phys. Rev. Lett., 116,
  191302

\bibitem[{{Quashnock} {et~al.}(1989){Quashnock}, {Loeb}, \&
  {Spergel}}]{Quashnocketal1989}
{Quashnock}, J.~M., {Loeb}, A., \& {Spergel}, D.~N. 1989, \apjl, 344, L49

\bibitem[{{Ratra}(1992)}]{Ratra1992}
{Ratra}, B. 1992, \apjl, 391, L1

\bibitem[{{Subramanian}(2016)}]{Subramanian2016}
{Subramanian}, K. 2016, Reports on Progress in Physics, 79, 076901

\bibitem[{{Tajima} {et~al.}(1992){Tajima}, {Cable}, {Shibata}, \&
  {Kulsrud}}]{Tajimaetal1992}
{Tajima}, T., {Cable}, S., {Shibata}, K., \& {Kulsrud}, R.~M. 1992, \apj, 390,
  309

\bibitem[{{Tevzadze} {et~al.}(2012){Tevzadze}, {Kisslinger}, {Brandenburg}, \&
  {Kahniashvili}}]{Tevzadzeetal2012}
{Tevzadze}, A.~G., {Kisslinger}, L., {Brandenburg}, A., \& {Kahniashvili}, T.
  2012, \apj, 759, 54

\bibitem[{{Trivedi} {et~al.}(2018){Trivedi}, {Reppin}, {Chluba}, \&
  {Banerjee}}]{Trivedi18}
{Trivedi}, P., {Reppin}, J., {Chluba}, J., \& {Banerjee}, R. 2018, \mnras, 481,
  3401

\bibitem[{Turner \& Widrow(1988)}]{TurnerWidrow1988}
Turner, M.~S. \& Widrow, L.~M. 1988, Phys. Rev. D, 37, 2743

\bibitem[{Vachaspati(2021)}]{Vachaspati:2020blt}
Vachaspati, T. 2021, Rept. Prog. Phys., 84, 074901

\bibitem[{Vazza {et~al.}(2017)Vazza, Brueggen, Gheller, Hackstein, Wittor, \&
  Hinz}]{va17cqg}
Vazza, F., Brueggen, M., Gheller, C., {et~al.} 2017, Classical and Quantum
  Gravity

\bibitem[{Vovk {et~al.}(2024)Vovk, Korochkin, Neronov, \&
  Semikoz}]{Vovk:2023qfk}
Vovk, I., Korochkin, A., Neronov, A., \& Semikoz, D. 2024, Astron. Astrophys.,
  683, A25

\bibitem[{{Xu} {et~al.}(2011){Xu}, {Li}, {Collins}, {Li}, \&
  {Norman}}]{Xuetal2011}
{Xu}, H., {Li}, H., {Collins}, D.~C., {Li}, S., \& {Norman}, M.~L. 2011, \apj,
  739, 77

\bibitem[{Zucca {et~al.}(2017)Zucca, Li, \& Pogosian}]{Zucca:2016iur}
Zucca, A., Li, Y., \& Pogosian, L. 2017, Phys. Rev. D, 95, 063506

\end{thebibliography}
\end{document}